\documentclass[conference]{IEEEtran}
\IEEEoverridecommandlockouts
\usepackage{cite}
\usepackage{amsmath,amssymb,amsfonts}
\usepackage{graphicx}
\usepackage{textcomp}
\usepackage{xcolor}
\usepackage{listings}
\usepackage{float}
\usepackage{lipsum}
\usepackage{lineno} 
\usepackage{algpseudocode} 
\usepackage[ruled,vlined]{algorithm2e}
\usepackage{epsfig}
\usepackage[export]{adjustbox}
\usepackage{multirow}
\usepackage{graphicx}
\usepackage{subfig}

\def\BibTeX{{\rm B\kern-.05em{\sc i\kern-.025em b}\kern-.08em
    T\kern-.1667em\lower.7ex\hbox{E}\kern-.125emX}}
\begin{document}

\title{Exploring the Impact of Serverless Computing on Peer To Peer Training Machine Learning}
\makeatletter
\newcommand{\linebreakand}{%
  \end{@IEEEauthorhalign}
  \hfill\mbox{}\par
  \mbox{}\hfill\begin{@IEEEauthorhalign}
}
\makeatother

\author{
\IEEEauthorblockN{
Amine BARRAK\IEEEauthorrefmark{1},
Ranim TRABELSI\IEEEauthorrefmark{1},
Fehmi JAAFAR\IEEEauthorrefmark{1},
Fabio PETRILLO\IEEEauthorrefmark{2}
}
\IEEEauthorblockA{\IEEEauthorrefmark{1}\textit{Department of Computer Science and Mathematics}, \textit{University of Quebec at Chicoutimi, UQAC}, Saguenay, Canada \\
Email: \{mabarrak, ranim.trabelsi1, fehmi.jaafar\}@uqac.ca}
\IEEEauthorblockA{\IEEEauthorrefmark{2}\textit{Département de génie logiciel}, \textit{École de Technologie Supérieure, ÉTS}, Montreal, QC \\
Email: fabio.petrillo@etsmtl.ca}
}

\newcommand{\al}{\textit{et al.}}
\newcommand{\cf}{{\textit{cf.,}}}
\newcommand{\aka}{{\textit{a.k.a.,}}}
\newcommand{\smallsection}[1]{\noindent\textbf{#1.}}
\newcommand{\eg}{{\textit{e.g.,}}}
\newcommand{\ie}{{\textit{i.e.,}}}

\maketitle

\begin{abstract}
\end{abstract}

The increasing demand for computational power in big data and machine learning has driven the development of distributed training methodologies. Among these, peer-to-peer (P2P) networks provide advantages such as enhanced scalability and fault tolerance. However, they also encounter challenges related to resource consumption, costs, and communication overhead as the number of participating peers grows. In this paper, we introduce a novel architecture that combines serverless computing with P2P networks for distributed training and present a method for efficient parallel gradient computation under resource constraints.

Our findings show a significant enhancement in gradient computation time, with up to a 97.34\% improvement compared to conventional P2P distributed training methods. As for costs, our examination confirmed that the serverless architecture could incur higher expenses, reaching up to 5.4 times more than instance-based architectures. It is essential to consider that these higher costs are associated with marked improvements in computation time, particularly under resource-constrained scenarios.

Despite the cost-time trade-off, the serverless approach still holds promise due to its pay-as-you-go model. Utilizing dynamic resource allocation, it enables faster training times and optimized resource utilization, making it a promising candidate for a wide range of machine learning applications.


\section{Introduction}
\label{sec:introduction} 

The exponential growth of data in the modern digital age  \cite{data2017exponential} has transformed the landscape of artificial intelligence (AI) and machine learning (ML), propelling these fields into a new era of innovation and discovery. This vast deluge of data, has given rise to increasingly sophisticated and complex models that can extract valuable insights and make accurate predictions \cite{verbraeken2020survey}. However, these sophisticated models pose a formidable challenge, due to the need for vast computational resources.

This escalating demand for computational power has led to the emergence of distributed training \cite{DistributedNN}. By harnessing the combined power of multiple devices, the training methodology encompasses the division of the dataset among a cohort of workers, each training their local model replicas in parallel and iteratively. To ensure convergence, the workers periodically synchronize their updated local models \cite{guerraoui2021garfield}.

Various topologies have been proposed in the literature \cite{verbraeken2020survey} to facilitate distributed training, including parameter server \cite{li2013parameter,P65,P76} and peer-to-peer architectures \cite{foster2003death,roy2019braintorrent,bellet2018personalized,wink2021approach,tang2020communication}. In the parameter server architecture, the worker nodes perform computations on their respective data partitions and communicate with the parameter server to update the global model. In contrast, peer-to-peer (P2P) architectures distribute the model parameters and computation across all nodes in the network, eliminating the need for a central coordinator\cite{verbraeken2020survey}.

Regardless of the topology employed for distributed training, developers often struggle with managing resources and navigating the complexities of ML training. This can result in over-provisioning and diminished productivity, posing challenges for ML users striving to achieve optimal outcomes \cite{P54}. 

To address these challenges, building machine learning (ML) on top of serverless computing platforms has emerged as an attractive solution that offers efficient resource management and scaling \cite{P09,P08,P46,P15}. By automatically scheduling stateless functions, serverless computing eliminates the need for developers to focus on infrastructure management
\cite{shafiei2022serverless, barrak}. However, ML systems are not inherently compatible with the Function-as-a-Service (FaaS) model due to limitations such as statelessness, lack of function-to-function communication, and restricted execution duration \cite{P54,P53}.

Numerous efforts have been made to optimize the utilization of FaaS platforms for managing ML pipelines\cite{P08,P51,P54,P39,sarroca2022mlless,P38,sampe2018serverless}. The implementation of parameter server architecture in a serverless environment demonstrated significant benefits, including reduced costs \cite{sarroca2022mlless}, scalability \cite{P54,P39}, and improved performance efficiency \cite{sampe2018serverless}.

Notwithstanding these encouraging findings, there remains a dearth of research elucidating the ramifications of serverless computing on peer-to-peer architecture. \textbf{To the best of our knowledge}, no research has been conducted to study the impact of serverless computing in a peer-to-peer environment.
 
Distributed training in peer-to-peer (P2P) networks offers benefits such as improved scalability and fault tolerance \cite{alqahtani2019performance}, but also presents challenges. As the network grows, communication, synchronization, and model update overheads increase, leading to latency and reduced training efficiency \cite{guerraoui2021garfield}. The diverse nature of devices in P2P networks can also cause imbalanced workloads and resource constraints, complicating the training process \cite{XIA2021100008}.

Another challenge faced during distributed training in P2P is the implementation of parallel batch processing inside each or the workers using popular machine learning frameworks like PyTorch. These frameworks often rely on the available and limited resources of individual workers to perform parallel computing on batches, which can lead to inefficiencies when resources are scarce \cite{kepner2018sparse, sattar2020data}. Consequently, these frameworks may resort to processing batches sequentially, which can result in longer training times and diminished performance.

In this paper, we present a novel approach to address all these challenges associated with distributed training in P2P networks by integrating serverless computing for parallel gradient computation. Our approach consists of the following components: (a) Incorporating serverless computing into the P2P training process, which eliminates the need to expand the number of workers in the network, effectively reducing communication and synchronization overhead and consequently enhancing training efficiency. (b) Introducing an advanced technique that leverages serverless functions and workflows for parallel gradient computation within each worker, ensuring efficient and accelerated gradient computation for each peer in the network, even in the presence of resource constraints.

Through a series of experiments and analyses \footnote{https://github.com/AmineBarrak/PeerToPeerServerless}, we demonstrate the effectiveness of our proposed approach in improving training, and optimizing resource utilization.

Our main contributions in this paper include:
\begin{itemize}
    \item \textit{Propose a novel architecture that integrates serverless computing into P2P networks for distributed training.}
    \item \textit{Introducing an advanced technique for efficient, parallel gradient computation within each peer, even under resource constraints.}
    \item \textit{Demonstrate the effectiveness of the proposed approach in improving training and optimizing resource utilization.}
\end{itemize}

\section{Background}
\label{sec:background}
In this background section, we delve into the intricate world of peer-to-peer machine learning. Additionally, we will explore the realm of serverless computing and workflow service state machines, such as AWS Step Functions.

\subsection{Peer To Peer architecture for distributed training}

Peer-to-Peer (P2P) architecture is a decentralized communication topology that is widely used in distributed systems. In a P2P system, nodes communicate directly with each other and there is no central point of control or coordination. 

In P2P training, the computational workload is distributed across multiple devices, creating a decentralized network where each device contributes its resources to collectively train the model. This approach can improve scalability, reduce training time, and minimize reliance on centralized infrastructure, making it a viable option for various applications, especially those with limited resources or rapidly changing workloads.

P2P training in machine learning presents an attractive alternative to traditional centralized training methods. By leveraging the distributed computing capabilities of multiple devices, P2P training can offer improved scalability, fault tolerance, and privacy preservation. However, challenges such as heterogeneity and resource constraints must be addressed to fully realize the potential of P2P training in machine learning applications.

\subsection{Serverless Computing}

Serverless computing is an emerging paradigm in cloud computing that enables developers to build and deploy applications without the need to manage server infrastructure. This model eliminates the need for developers to worry about infrastructure scaling, server maintenance, and other low-level tasks, allowing them to focus on creating business logic.

Serverless computing is built on the concept of Function-as-a-Service (FaaS), which provides developers with a platform to deploy and run small pieces of code, called functions, in response to events. When an event triggers a function, the cloud provider provisions the necessary infrastructure to run the function, and then releases it once the function completes its execution.

The benefits of serverless computing, such as cost-effectiveness, scalability, flexibility, and ease of use, make it a promising approach for machine learning applications, enabling efficient resource management and rapid model development.

\subsection{Serverless AWS Step Function Workflow}

The AWS Step Function Workflow enables developers to design, execute, and monitor multi-step workflows, addressing the complexity of manually managing multiple serverless functions (e.g., AWS Lambda Function \cite{Serverle24:online}). By defining a state machine using the Amazon States Language, developers can create long-running workflows that are easy to understand and maintain, improving the overall coordination of serverless applications.

\section{Methodology and System Design}
\label{sec:study}

In this section, we present a novel P2P training ML system based on Serverless computing, focusing on the design architecture, algorithm, and techniques to reduce peer overload. Our approach aims to improve efficiency, scalability, and alleviate resource constraints in ML training.

\subsection{Design Architecture of Peer to Peer training Machine Learning based on Serverless Computing}
\begin{figure*}[!h]
 \centering
 \includegraphics[scale=0.5]{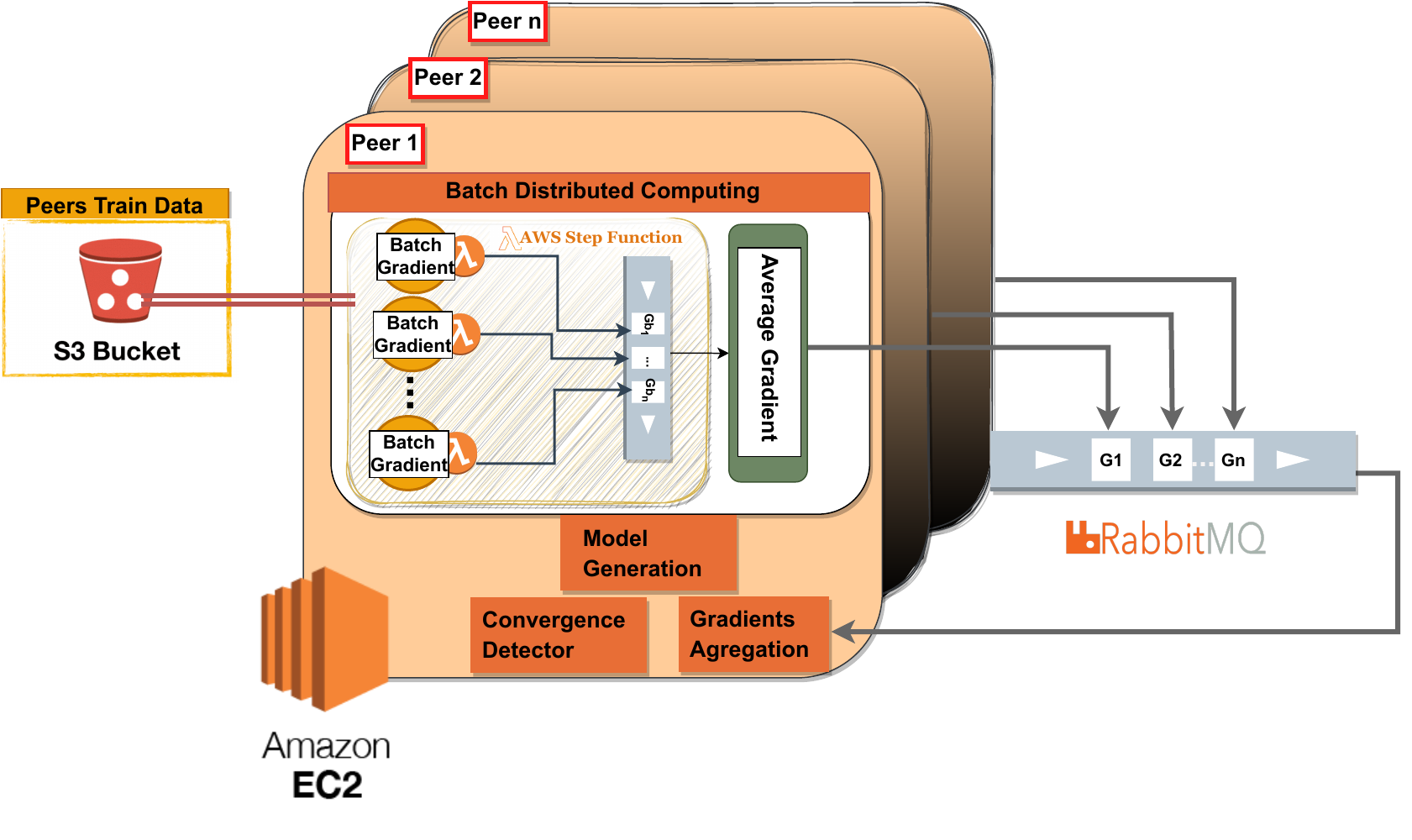}
 \caption{Overview of the proposed Peer To Peer training based on Serverless computing}
 \label{fig:arch}
\end{figure*}
We implemented our approach using AWS Lambda due to its 15-minute timeout and 10GB RAM availability \cite{Lambdaqu81:online}. Comparable services exist on platforms like Google Cloud Functions, Azure Functions, and IBM Cloud Functions.

Figure \ref{fig:arch} describes the overall proposed architecture. 
During the training of deep learning models, PyTorch strives to maximize resource utilization efficiently. However, ML frameworks \ie PyTorch, do not inherently possess a mechanism to seamlessly transition between parallel and sequential processing under resource constraints. In real-world scenarios, ML frameworks leverage a GPU for computations when available and default to the CPU when GPU resources are not accessible.

By harnessing the power of serverless computing, our system architecture enables parallel gradient computations across multiple Lambda functions, leading to a substantial reduction in overall computation time. We thoroughly examine the intricacies of our peer-to-peer architecture, which consists of four integral system components. An overview of the peer to peer ML system based on Serverless computing architecture is depicted in Figure \ref{fig:arch}.

\textbf{AWS S3 Buckets} : 
In a peer-to-peer network, data is systematically partitioned into discrete segments, with each peer's assigned portion subsequently uploaded to a dedicated S3 bucket. This approach guarantees seamless access to their own data for each peer, while simultaneously leveraging the high-performance, cloud-based architecture of S3.

\textbf{AWS Lambda Function} : 
We strategically chose to implement parallel batch processing, a complex task made feasible by employing AWS Lambda serverless functions. By harnessing AWS Lambda's capabilities, we link each data batch to a specific Lambda function responsible for executing the necessary gradients computations. This approach significantly reduces total computation time through the wise distribution of workloads across multiple Lambda function instances, accelerating data processing and cutting down the time needed to complete processing the training set. Additionally, We integrate AWS Step Functions to manage, orchestrate and invoke the Lambda serverless parallel computing process, adapting to the availability of data batches and ensuring efficient handling of the workload.

\textbf{EC2 Instance}: 
Each EC2 instance in our system architecture, assigned to individual peers, carries multiple responsibilities. First, it acts as a trigger for invoking Lambda functions responsible for essential gradient computation. Additionally, it includes a crucial set of features that enable gradient exchange between peers. Ultimately, the EC2 instance is equipped with a specialized feature to detect model convergence, further boosting the overall efficiency of the system.

\textbf{RabbitMQ} : 
The proposed architecture relies on the utilization of RabbitMQ, that enable seamless communication between peers. After computing gradient averages over batches, a peer publishes the resultant data to its dedicated queue. Other peers in the network can access the gradients published in the queue, enabling efficient and seamless information sharing. This is a critical aspect of our methodology, as it allows each peer to access the required information quickly and accurately to perform computations. RabbitMQ's reliability and security ensure smooth and secure data transmission and communication, promoting an efficient processing of complex data sets.

\subsection{Peer to Peer training Machine Learning}

\begin{algorithm}
\DontPrintSemicolon
\caption{P2P ML Distributed Training}
\label{algo_training}
\KwIn{Deep Neural Network training workload with its \textbf{input train dataset $D$ with size $n$}, \textbf{validation dataset $V$}, the \textbf{number of peers $P$}, \textbf{model size $d_m$}, the \textbf{learning rate $\eta$}, the \textbf{batch size $B$}, and the \textbf{number of epochs $E$}.}

\KwOut{The trained model with updated weights $\theta^*$. }
\underline{\textbf{Peer of rank} $r = 0, \dots, P:$}\newline
$\triangleright$ Each peer simultaneously implements: \; 
Initialize: $Gradients\_Peers$=\{\}\; 
Initialize communication channels in RabbitMQ\;
Initialize a dedicated peer queue $q_r$\;

Initialize model parameters randomly as $\theta_0\in \mathbb{R}^d$\;
\For{epoch e = 1 \textbf{to} $E$} {
    Load a unique partition of data $D_r$\;
    Randomly partition the subset $D_r$ into $m$
    batches of size $B$\;
    \For{each batch $b$} {
        $g_{t,b}\leftarrow$ \textbf{ComputeBatchGradients}$(\theta_{t-1})$\; 
        }
        \textbf{AverageBatchesGradients} as  $g_{t,r}\leftarrow \frac{1}{m} \sum_{b=1}^{m} g_{t,b}$\; 
        
        $Gradients\_Peers[r]\leftarrow g_{t,r}$ \; 
        \textbf{SendGradientsToMyQueue}$(g_{t,r},q_{r})$\;
        \For{i from 0 to P}{
            \If{i is not equal to r}{
                $g_{t,i}\leftarrow$\textbf{ConsumeGradientsFromQueue}$(q_{i})$\;
                WaitUntilReceptionDone()\;
                $Gradients\_Peers[i] \leftarrow g_{t,i}$\;
            }
        }
        $g_{t}\leftarrow$ \textbf {AverageGradients}($Gradients\_Peers$)\newline 
        \If{is\_synchronous}{
            \textbf{SynchronisationBarrier()}
        } 

    Update the model as $\theta_t \gets \theta_{t-1} + \eta \cdot g_t$\;
    
    \If {\textbf{DetectConvergence($\theta_{t}$,$V$)}}
    {
        \textbf{Return} updated model $\theta^*$.
    }
   
}

\end{algorithm}

We specify a peer-to-peer architecture that leverages distributed computation for the purpose of training machine learning models. Algorithm \ref{algo_training} present the logic we followed. Initially, a workload is provided that includes the Deep Neural Network (DNN) model and the training dataset, along with parameters specifying the number of peers (P), batch size (B), and training epochs (E).

Additionally, each peer has an array of key-value pairs, where the key is the peer's rank (ID) and the value is the computed gradient.

We explain in the following the different sections of the algorithm.

\subsubsection{Dataset Preprocessing}
Within our system architecture, we have integrated a preprocessing stage to to transform the training dataset using methods like min-max scaling, standardization, and normalization. After preprocessing, the dataset is divided into partitions for each peer in the training process. A dataloader is implemented to further split the partitions into batches, which are then stored in designated Amazon S3 cloud storage buckets.

\subsubsection{Compute Batch Gradients} The peer-to-peer training paradigm entails a multi-stage process wherein each worker subdivides its designated data subset into smaller batches, which are intended to expedite the training and convergence process by allowing each worker to compute gradients for smaller subsets of the data. During the training phase, each worker calculates the gradients for the batches of data it has processed and subsequently averages these gradients across all batches. This crucial step enables each worker to obtain an accurate representation of the gradients for its designated subset of data.

\subsubsection{Communication Protocol} 
To communicate between peers, we used Amazon MQ's RabbitMQ for exchanging gradients between multiple peers during the model synchronization process. Each peer is assigned a dedicated queue that contains a single, persistent gradient message. When a new gradient is generated, it replaces the previous one in the queue, ensuring that the latest gradient is always available for consumption by other peers.

Peers can access and consume gradient messages from all other queues without deleting them, which promotes efficient gradient exchange and prevents data loss in case of temporary disruptions. The persistence of gradient messages guarantees the availability of the necessary information for model synchronization, even under challenging network conditions.

When peers are ready to synchronize their models, they read the gradient messages from all other queues, excluding their own. This process allows them to effectively update their models based on the gradients received from other peers, streamlining the distributed training process across the entire system.

To store received gradients from peers, a dictionary is created, where the peer's rank serves as the key to map to its corresponding received gradient.
Each peer retains the received gradients in the local dictionary, and if the dictionary's size exceeds a threshold predefined in advance, the peer retrieves the gradients and calculates their average. The worker then updates its model parameters in accordance with the result. This iterative process continues for a predetermined number of epochs, as established by the input hyperparameters.

To overcome Amazon MQ's message size limitations (100MB per message), large files are stored in Amazon S3 and referenced using UUIDs. Sending UUIDs through Amazon MQ enables efficient, scalable data transfer without compromising performance or reliability, providing a flexible solution for seamless data exchange.

\subsubsection{Compression / Decompression}

we address the challenge of high communication overhead by incorporating the QSGD algorithm \cite{alistarh2017qsgd}. This algorithm uses a compression technique to quantize gradients before transmission, reducing the size of transmitted gradients and leading to improved training efficiency.

\subsubsection{Average Gradients}
After receiving gradients from other peers, each peer aggregates the gradients by computing their average and uses this averaged gradient to update their local model parameters. The advantage of this approach is that it allows each peer to learn from the gradients computed by other peers, resulting in a more accurate representation of the global gradients.

\subsubsection{Synchronous \& Asynchronous Gradient Computation}

In the following stage, the worker simultaneously distributes the averaged gradients to all other workers in the network and receives from them their averaged gradients as well. This process can be executed using either synchronous or asynchronous approaches. Figure \ref{fig:sync_async} show an example of synchronous and asynchronous communication using four workers.

\begin{figure}[H]
 \centering
 \includegraphics[scale=0.55]{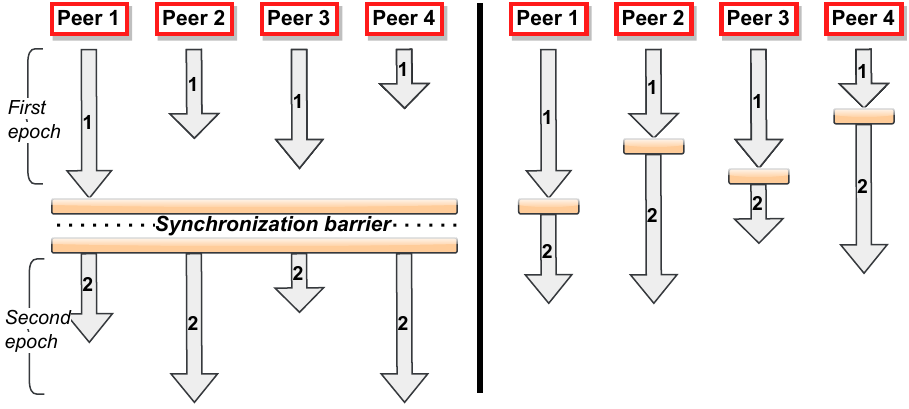}
 \caption{Synchronous(left) and asynchronous(righ) Communication}
 \label{fig:sync_async}
\end{figure}
 
\textbf{In the asynchronous communication}, Amazon MQ's RabbitMQ service provide a separate dedicated queues for each peer. These queues store the latest gradients generated by each peer, and they can be accessed and consumed by other peers without having to wait for every peer to finish their gradient computation. This means that a peer can start updating its model with the latest available gradients from other peers, without waiting for gradients from slower peers or those experiencing temporary disruptions.

\textbf{In the synchronous communication,} a synchronization barrier is added to ensure that all peers progress through the distributed training process together.

Synchronizing autonomous peers in a distributed system is challenging, especially when using RabbitMQ queues for gradient communication. Factors like varying resource availability can cause some peers to progress through epochs at different speeds. To address this issue, we have implemented a RabbitMQ-based synchronization mechanism. Each peer sends a message to a designated synchronization queue, signifying the completion of gradient computation, sending, and receiving for all connected peers. Once the size of this synchronization queue matches the total number of peers, it indicates that all peers have completed the current epochs, and they can then proceed to the next one in a coordinated manner.

\subsubsection{Convergence Detection}
To detect model convergence, two key techniques are used: \textit{ReduceLROnPlateau} and \textit{Early Stopping}. ReduceLROnPlateau adjusts the learning rate during training, improving generalization by preventing overshooting the loss function's minimum. It monitors model performance on a validation dataset, reducing the learning rate if improvement stalls.

Early stopping detects convergence by tracking performance during training and stopping when performance degrades, preventing overfitting. If convergence isn't reached through these techniques, the epoch limit determines the maximum training iterations.

Achieving convergence ensures the model's accuracy and effectiveness in making predictions. 

\subsubsection{Memory, CPU and Time metrics collection:}
To assess and diagnose the efficiency of the system architecture, several Python libraries are used for recording performance metrics. Tracemalloc is utilized for measuring RAM utilization, psutil for monitoring CPU usage in real-time, and the perf\_counter function for evaluating time-based performance. These tools enable a deep understanding of system performance and identification of areas requiring optimization or improvement.

\subsection{Serverless to reduce a Peer overload computing:}

We leverage AWS Lambda for serverless parallel batch processing, enabling efficient workload distribution and reducing computation time. By assigning specific Lambda functions to data batches, we effectively manage gradients computations. AWS Step Functions orchestrate the Lambda functions, adapting to data batch availability for optimal workload handling. This serverless approach minimizes peer overload and accelerates training set processing, enhancing overall performance.

\section{Experimental Setup}
\label{sec:setup}

This section details our experimental setup to evaluate the performance of various CNN models across different datasets on the proposed architectures.

\subsection{Datasets}
\textbf{MNIST:} The MNIST Handwritten Digit Collection \cite{deng2012mnist} consists of 60,000 samples of handwritten numerals, each categorized into one of ten classes.

\textbf{CIFAR:} The CIFAR Image Dataset \cite{krizhevsky2010convolutional} encompasses 60,000 color images spanning ten distinct classes, such as automobiles, animals, and objects. Each category contains 6,000 images that are evenly distributed.

\subsection{Model Architectures and Hyperparameters}
\textbf{SqueezeNet 1.1:} SqueezeNet 1.1 is an efficient CNN architecture \cite{iandola2016squeezenet}, with fewer parameters (~1.2 million) and a small model size ($<$5MB).

\textbf{MobileNet V3 Small:} MobileNet V3 Small \cite{koonce2021mobilenetv3} is a lightweight CNN tailored for mobile and edge devices, featuring inverted residual blocks, linear bottlenecks, and squeeze-and-excitation modules. With approximately 2.5 million trainable parameters and a compact model size,

\textbf{VGG-11:} VGG-11 is a deep convolutional neural network (CNN) architecture developed for image classification tasks \cite{simonyan2014very}. It is a variation of the VGG family, with 11 weight layers, including convolutional and fully connected layers. With an input resolution of 224x224 and approximately 132.9 million trainable parameters. 

\subsection{EC2 Instances configuration for peers}

We aim to determine the ideal machine instance for three different neural network models: Vgg11, MobileNet V3 Small, and SqueezeNet 1.1. We started with the smallest available machine instance and trained the models on it. If the machine crashed due to resource limitations during training, we moved up to the next larger machine instance until we found one that was able to train the model without issue. Additionally, we incrementally increased the number of peers during the experimentation, starting with 4 and adding 4 peers at a time until we reached 12 peers, to determine the computation and communication resources usage. Ultimately, we determined that the Vgg11 model require t2.large instance, while the MobileNet V3 Small and SqueezeNet 1.1 models could be trained model could be trained on t2.medium instance. This approach allowed us to optimize the use of resources and achieve optimal performance for each model, taking into account both computation and cost.

\subsection{Serverless client functions configuration}
In the following, we discuss our approach to implementing a serverless training workflow by leveraging AWS Step Functions and Lambda functions for parallel gradient computation and batch processing.

\subsubsection{Serverless AWS Lambda Configuration for Gradient Computation}
We prepared an AWS lambda serverless function for machine learning batch training. The function is designed to be invoked with essential parameters such as the specific model, batch identifier, optimizer, learning rate, and loss function. To obtain the necessary data batch for training, the function accesses an S3 bucket, where we have pre-processed and stored batches.

To facilitate seamless deployment on our custom ARM architecture, we packaged the machine learning dependencies, including the Pytorch library, in a zip file with a size less than 50MB. If additional dependencies are needed, they can be incorporated as separate layers within the AWS Lambda service. This approach allows for a modular structure while complying with the service's constraints. The total size of the unzipped files must not exceed 250MB, ensuring that the serverless function remains within the allowable resource limits, ultimately fostering efficient and scalable training processes in our custom ARM-based environment.

\subsubsection{Serverless AWS Lambda Pricing}
One of the key factors in pricing for AWS Lambda is the amount of memory allocated to the function. The prices of AWS Lambda are calculated based on the amount of memory allocated to the function and the duration of the execution.

The objective is to compare the costs of running the same workload using EC2 peer to peer instances without serverless and using EC2 small instances by invoking serverless lambda for parallel compute gradients. This comparison will give us insights into the cost-effectiveness of using serverless computing in contrast to traditional computing methods. 

\subsubsection{Dynamic AWS Step Function State Machine for Parallel Batch Processing}

We have developed a Dynamic State Machine using AWS Step Functions, designed to compute parallel batch gradients on serverless Lambda functions. This state machine is generated dynamically according to the given batch number, allowing it to accommodate varying batch sizes. By leveraging the parallel computing capabilities of AWS Step Functions, each Lambda invocation processes an assigned batch saved in an S3 bucket. Once the state machine is deployed, it is invoked with the necessary input, which includes the total number of batches and the data required for the Lambda function to compute gradients corresponding to each data batch. This data encompasses the model, batch, optimizer, learning rate, and loss function. Our approach effectively enables parallel processing of gradient computations within a serverless environment using AWS Step Functions and Lambda functions. 

\section{Experimental Results}
\label{sec:results}
 \setlength{\belowcaptionskip}{-10pt}
\begin{figure*}[t]
    \centering
    \subfloat[Four Peers\label{fig:parameter-serverq}]{{\includegraphics[width={0.3\textwidth}]{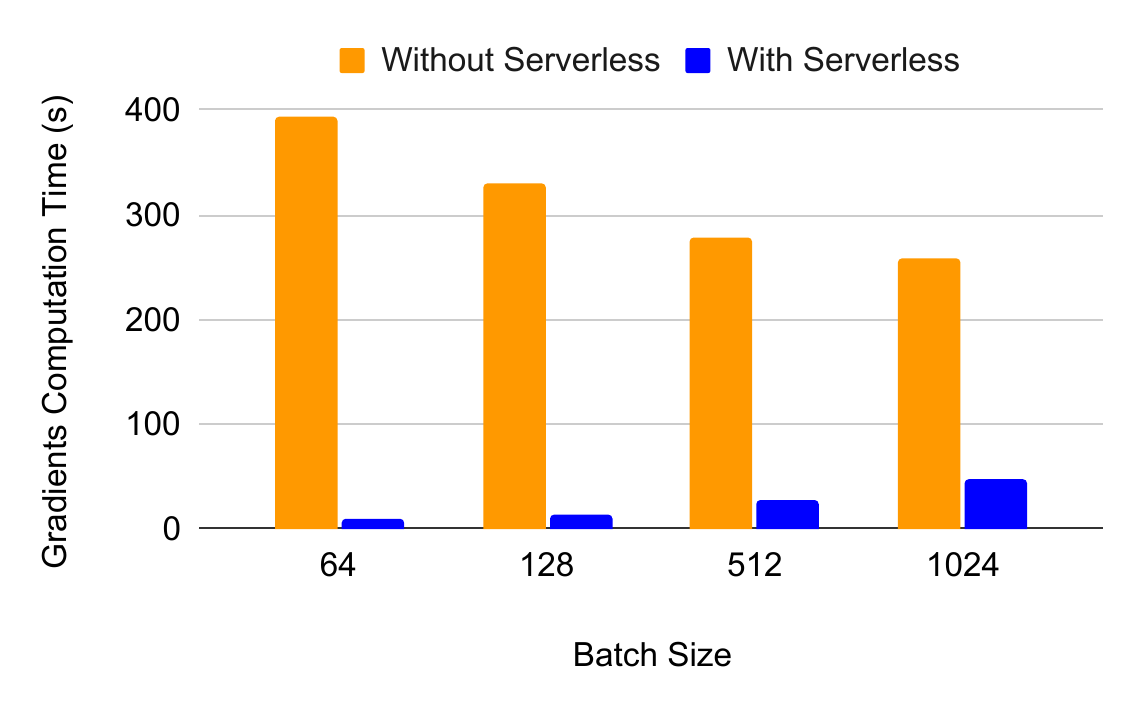} }}%
    \qquad
    \subfloat[Eight Peers\label{fig:p2p1}]{{\includegraphics[width={0.3\textwidth}]{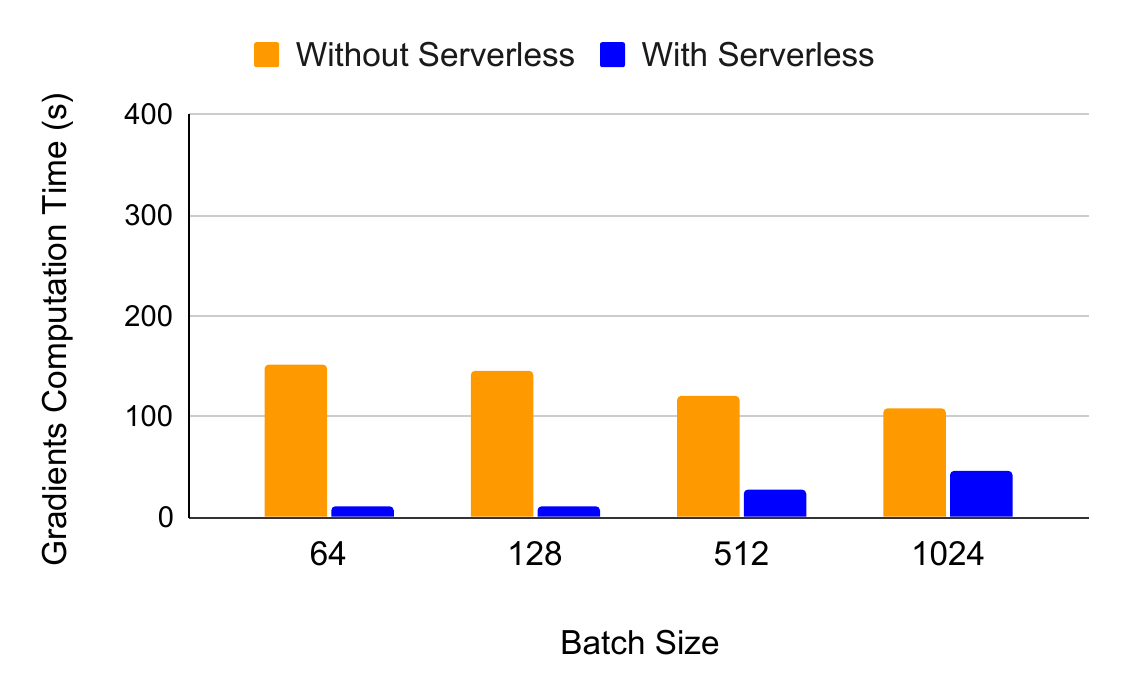} }}%
    \qquad
    \subfloat[Twelve Peers\label{fig:p2p1}]{{\includegraphics[width={0.3\textwidth}]{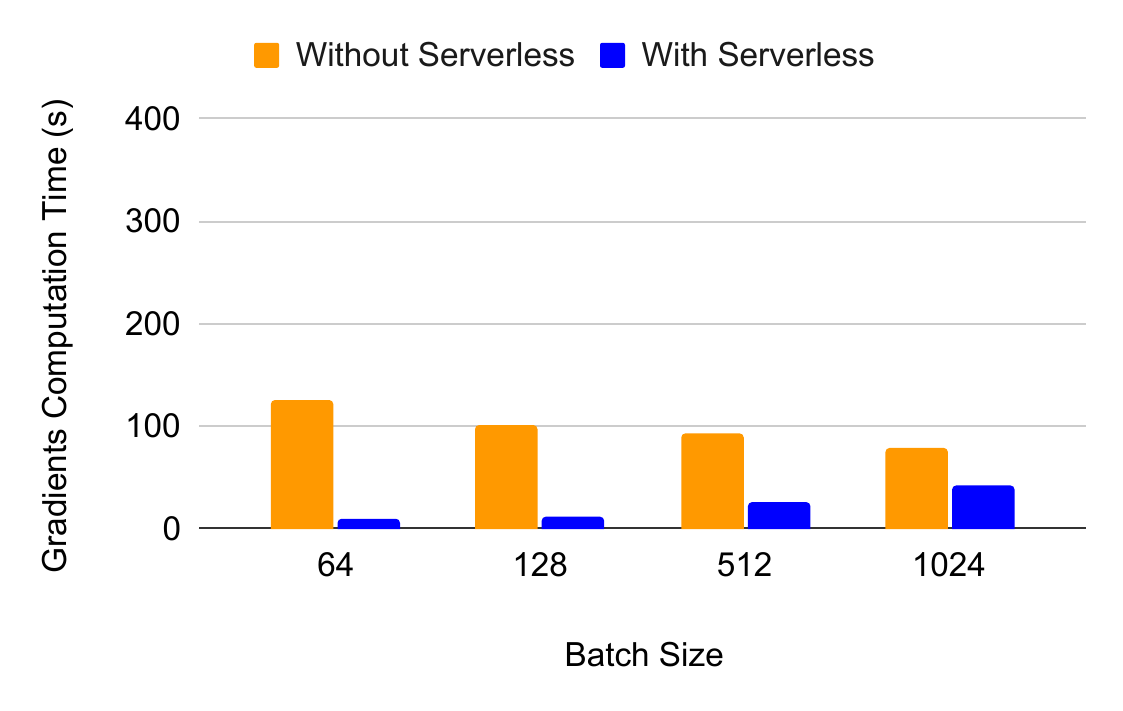} }}%
    \caption{Comparison of Processing Training Time on Gradients Computing for different number of Peers and batch sizes in Peer to Peer Training with and Without Serverless}%
    \label{fig:time_peers}%
\end{figure*}
In this section, we present the results of our experiments, focusing on distributed deep learning aspects including resource requirements, serverless efficiency, communication overhead, and synchronization barriers in peer-to-peer training.

\subsection{Identify tasks needing expensive computational level}

To determine the resource usage and identify computationally expensive tasks in a distributed peer-to-peer training setup. In this setup, four worker nodes collaborate to train a machine learning model. The experiment focuses on measuring the resource usage at different stages of the distributed training process, including computing gradients, sending gradients, receiving gradients, updating the model, and convergence detection, is monitored and captured.

Metrics such as CPU usage, memory consumption, and processing time are recorded for each stage. The experiment continues to four epochs and the average per epoch is computed. Afterward, we compare resource consumption across different stages and identify the most computationally demanding tasks.

\begin{table}[h]
\caption{Evaluating Resource Usage in Distributed Peer-to-Peer Training with Four Workers and 30 Batches; Goal is to Determine the Most Resource Consuming in Term of CPU/Memory and Processing Time}
\centering
\begin{adjustbox}{max width=9cm}
\begin{tabular}{|l|c|l|l|l|l|l|l|}
\hline
\textbf{\begin{tabular}[c]{@{}l@{}}Model \\ (instance type)\end{tabular}} & \multicolumn{1}{l|}{\textbf{Dataset}} & \textbf{\begin{tabular}[c]{@{}l@{}}Training\\  Stage\end{tabular}} & \textbf{\begin{tabular}[c]{@{}l@{}}Compute \\ Gradients\\ (per batch)\end{tabular}} & \textbf{\begin{tabular}[c]{@{}l@{}}Send \\ Gradients\end{tabular}} & \textbf{\begin{tabular}[c]{@{}l@{}}Receive \\ Gradients\end{tabular}} & \textbf{\begin{tabular}[c]{@{}l@{}}Model \\ Update\end{tabular}} & \textbf{\begin{tabular}[c]{@{}l@{}}Convergence \\ detection\end{tabular}} \\ \hline
\multirow{3}{*}{\textbf{\begin{tabular}[c]{@{}l@{}}squeezenet 1.1 \\ (t2.medium)\end{tabular}}} & \multirow{3}{*}{\begin{tabular}[c]{@{}c@{}}MNIST\\  / CIFAR\end{tabular}} & \begin{tabular}[c]{@{}l@{}}CPU \\ Usage (\%)\end{tabular} & 194,82 / 195,45 & 39,37 / 43,95 & 71,85 / 73,72 & 122,4 / 141,5 & 198,17 /  196,17 \\ \cline{3-8} 
 &  & \begin{tabular}[c]{@{}l@{}}Memory \\ (MB)\end{tabular} & 600 / 570 & 568,29 / 566,64 & 555,32 / 569,91 & 566 / 530 & 574 / 540 \\ \cline{3-8} 
 &  & \begin{tabular}[c]{@{}l@{}}Processing \\ Time (s)\end{tabular} & 14,93 / 14,01 & 0,084 / 0,08 & 0,25 / 0,27 & 0,18 /  0,0052 & 0,19 / 0,16 \\ \hline
\multirow{3}{*}{\textbf{\begin{tabular}[c]{@{}l@{}}MobileNet \\ V3 Small \\ (t2.medium)\end{tabular}}} & \multirow{3}{*}{\begin{tabular}[c]{@{}c@{}}MNIST\\  / CIFAR\end{tabular}} & \begin{tabular}[c]{@{}l@{}}CPU \\ Usage (\%)\end{tabular} & 197,87 / 198 & 40,45 / 40,92 & 75,3 / 81,42 & 147,6 / 134,8 & 198,05 / 198 \\ \cline{3-8} 
 &  & \begin{tabular}[c]{@{}l@{}}Memory \\ (MB)\end{tabular} & 840 / 640 & 835,70 / 630,86 & 843,81 / 624,33 & 786 / 780 & 800 / 800 \\ \cline{3-8} 
 &  & \begin{tabular}[c]{@{}l@{}}Processing \\ Time (s)\end{tabular} & 29,72 / 24,01 & 0,11 / 0,11 & 0,38 / 0,43 & 0,015 / 0,016 & 1,12 / 0,92 \\ \hline
\multicolumn{1}{|c|}{\multirow{3}{*}{\textbf{\begin{tabular}[c]{@{}c@{}}VGG 11\\ ( t2.large )\end{tabular}}}} & \multirow{3}{*}{\begin{tabular}[c]{@{}c@{}}MNIST\\  / CIFAR\end{tabular}} & \begin{tabular}[c]{@{}l@{}}CPU \\ Usage (\%)\end{tabular} & 198,4 / 198,2 & 65,22 / 74 & 53,72 / 56 & 166 / 154 & 198,4 / 198,8 \\ \cline{3-8} 
\multicolumn{1}{|c|}{} &  & \begin{tabular}[c]{@{}l@{}}Memory \\ (GB)\end{tabular} & 4,10 / 4,24 & 3,19 / 3,075 & 4,8 / 4,33 & 2,4 / 2,46 & 2,41 / 2,4 \\ \cline{3-8} 
\multicolumn{1}{|c|}{} &  & \begin{tabular}[c]{@{}l@{}}Processing \\ Time (s)\end{tabular} & 104,37 / 104,20 & 7,38 / 6,72 & 15,55 / 19,54 & 4,8 / 4,2 & 9,20/7,6 \\ \hline
\end{tabular}
\end{adjustbox}
\label{tab:evaluation}
\end{table}

According to the results of our experimental investigation on three different models, namely VGG11, MobileNetV3 Small, and SqueezeNet, using two distinct datasets, MNIST and CIFAR, we have identified the most resource-intensive step during the training process. As demonstrated in the tabulated data of Table \ref{tab:evaluation}, the computation of gradients consumes a substantial amount of computational resources and memory, particularly for VGG11, which requires approximately 4 GB of memory per batch, and given that we executed 30 batches during our experiment. In comparison to other stages, such as sending and receiving data, updating models, and detecting convergence, the computation of gradients resulted in the highest CPU usage. As a result, it is reasonable to recommend the migration of the computation of gradients to a serverless infrastructure, which can reduce the overheads associated with managing and provisioning resources.

\subsection{Evaluation of Serverless Infrastructure for Gradient Computing}

Throughout this section, we conducted a series of experiments to evaluate the impact of serverless infrastructure on the performance and cost of gradients computing. We evaluate two distinct architectures to assess the impact of serverless integration on resource utilization and cost. In the first architecture, we train a VGG11 model and MNIST dataset with \textit{t2.large} instances. In the second architecture, we train the same model with \textit{t2.small} instances, while offloading high-computational tasks to a distributed lambda serverless infrastructure. 

\subsubsection{Computation Time Comparison: Serverless vs. Instance-based Architectures for Gradient Computing} 

We examined different architectures, batch sizes, and numbers of workers to gain a comprehensive understanding of the potential benefits and challenges associated with serverless integration in terms of execution time of the gradients computation.
The findings from our experiments are illustrated in a bar plot figure\ref{fig:time_peers} , where we have two bars for each batch size – one representing the time taken with serverless infrastructure (blue bar) and the other without serverless infrastructure (orange bar). This visual representation clearly highlights the significant improvements in the time taken to compute batches when employing serverless infrastructure across various batch sizes (64, 128, 512, and 1024) and numbers of workers (4, 8, and 12). For instance, in a configuration with 4 workers and a batch size of 64, the blue bar (serverless) is considerably shorter than the orange bar (non-serverless), demonstrating a remarkable 97.34\% reduction in the time taken to compute batches. Similarly, with 8 workers and a batch size of 128, the improvement reaches 92.04\%. However, it is worth noting that the improvement tends to decrease as the number of workers increases, especially for larger batch sizes.

\subsubsection{Cost Comparison: Serverless vs. Instance-based Architectures for Gradient Computing}

 \setlength{\belowcaptionskip}{-10pt}
\begin{figure*}[t]
    \centering
    \subfloat[VGG\label{fig:time_gradients}]{{\includegraphics[width={0.37\textwidth}]{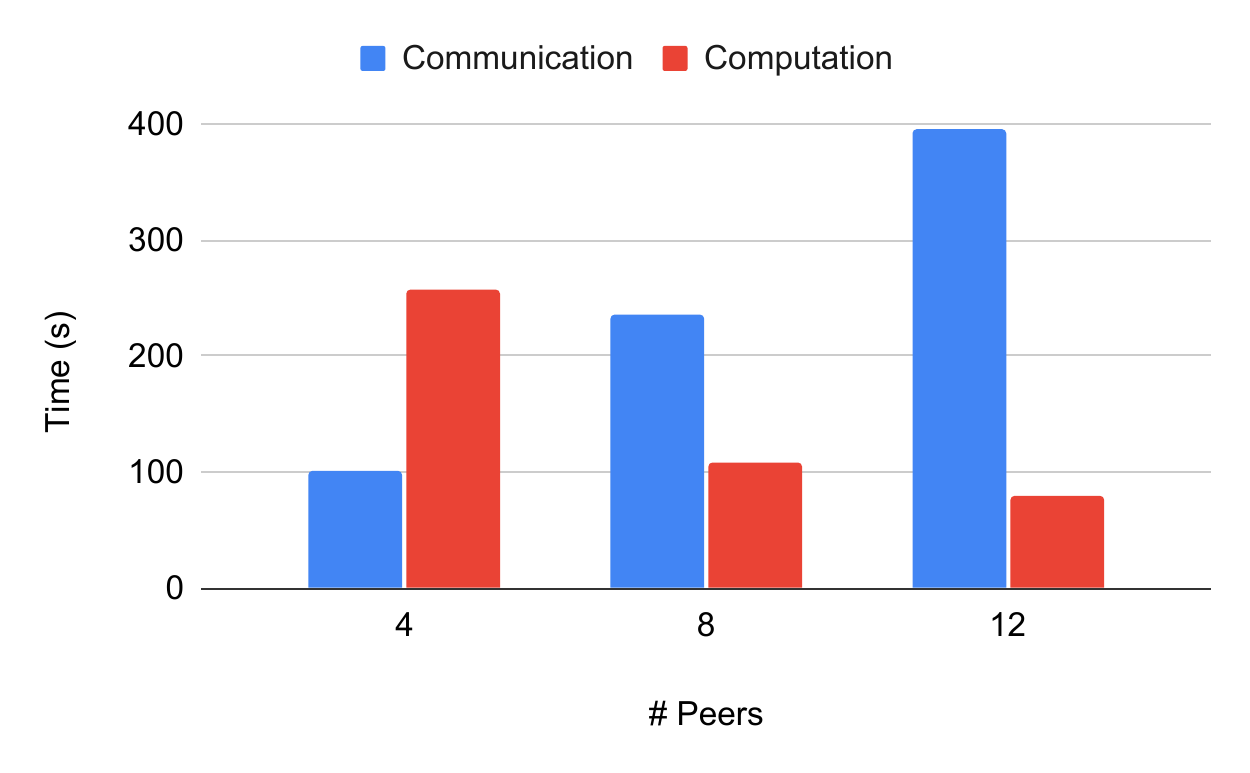} }}%
    \qquad
    \subfloat[MobileNet V3 Small\label{fig:cost}]{{\includegraphics[width={0.37\textwidth}]{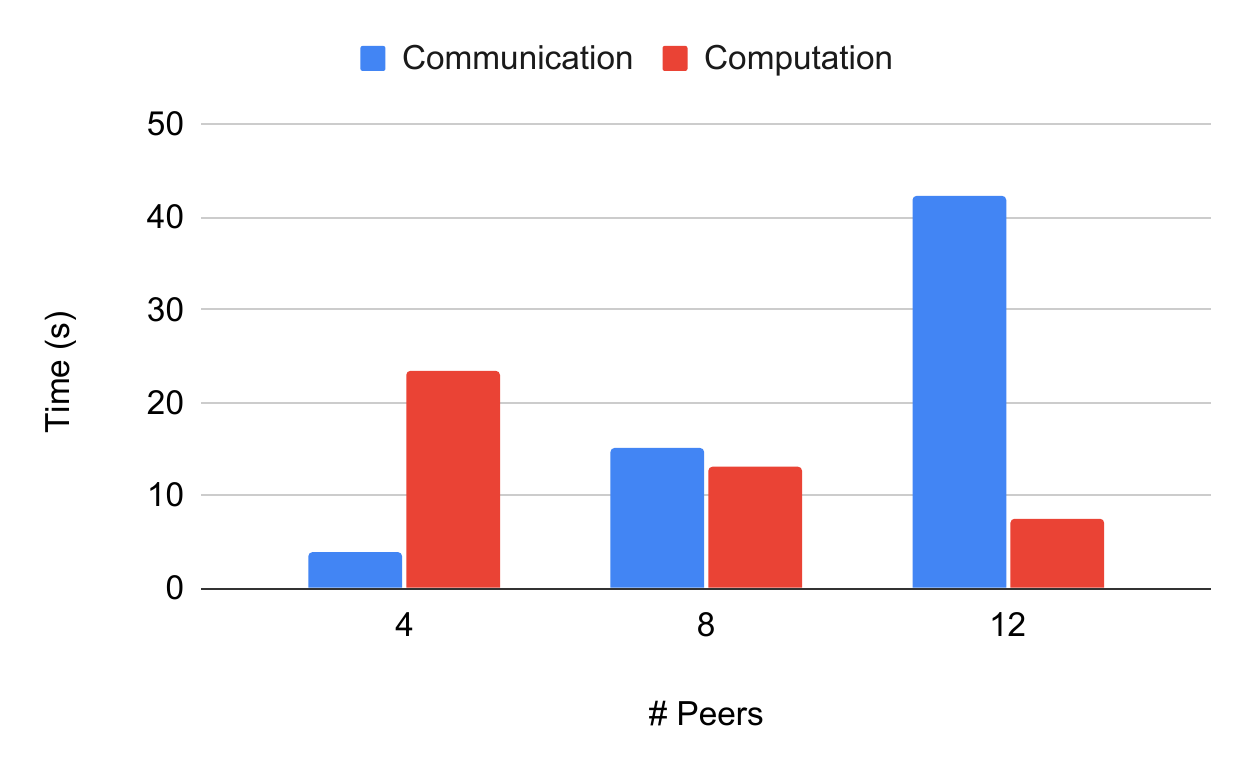} }}%
    \caption{Gradients Computation and communication time per \# Peers on VGG11 and MobileNet V3 Small (1024 batch size)}%
    \label{fig:comm}%

\end{figure*}
In the previous experiment, we evaluated the impact of serverless infrastructure on computation time for gradient computing in peer-to-peer training. The results demonstrated significant improvements across varying batch sizes and numbers of workers, especially with a four-worker setup. This finding prompted us to delve deeper into the cost analysis for this scenario.

In this section, we present a cost comparison between serverless and instance-based architectures for gradient computing, focusing on a case study involving four workers, the VGG11 model, and the MNIST dataset. Tables \ref{tab:with_serverless} and \ref{tab:without_serverless} detail the time and cost evaluation for computing gradients with different batch sizes in both architectural scenarios. Lambda memory size was set to match the minimal functional requirements for gradient computation.

In our cost comparison analysis, the estimated cost per peer was calculated as follows:

\begin{align}
\text{Cost per Peer}_{\text{serverless}} & =  \textbf{[} \text{Lambda Cost} \times \text{Num of batches} \nonumber \\
& \quad + \text{EC2 Cost}\textbf{]} \nonumber \\
& \quad \times \text{Computation Time}
\label{eq:serverless_cost}
\end{align}

\begin{align}
\text{Cost per Peer}_{\text{instance-based}} & = \text{EC2 Cost} \times \text{Computation Time}
\label{eq:instance_based_cost}
\end{align}

From Table \ref{tab:with_serverless}, we observe that for serverless architecture, as the batch size decreases, so does the computation time, leading to variable costs per batch size. However, the number of batches, also increases, affecting the lambda costs since each batch is a separate invocation of the lambda function. Hence, while larger batch sizes increase efficiency in computation time, they also necessitate more resources, thus increasing the costs.

In comparison, Table \ref{tab:without_serverless} presents the costs associated with the instance-based architecture, showing a clear increase in the costs as the batch size decreases. The cost differences between the two architectures can be attributed to the use of different instance types (t2-small, t2.large) and the varying memory size requirements for the lambda functions in the serverless architecture.

For a detailed understanding of the cost dynamics, we scrutinized the estimated cost of computing gradients (in USD) for both architectures across all batch sizes. For a batch size of 1024, we found that the serverless architecture costs approximately 5.34 times more than the instance-based architecture. However, this discrepancy in cost decreases with smaller batch sizes.

The results highlight a greater cost when utilizing a serverless architecture with low resource instances, it's important to consider the time-efficiency gains. As there is a trade-off between the significant improvements in computation time and the cost associated with using serverless infrastructure. It is essential for researchers and practitioners to consider their specific requirements, such as training time constraints and budget limitations, when selecting an architecture for gradient computing.

\begin{table}[H]
\caption{Time and Cost Evaluation of Compute Gradients in Peer to Peer Training with Serverless; Model trained on VGG11, MNIST dataset, and Four Peers}
\centering
\begin{adjustbox}{max width=9cm}
\begin{tabular}{|c|c|c|c|c|}
\hline
Batch Size                   & 1024       & 512        & 128        & 64         \\ \hline
Number of batches            & 15         & 30         & 118        & 235        \\\hline
Instance Type                & t2-small   & t2-small   & t2-small   & t2-small   \\\hline
\begin{tabular}[c]{@{}c@{}}Lambda \\ Memory size\end{tabular}  & 4400 MB    & 2800 MB    & 1800 MB    & 1700 MB    \\\hline
\begin{tabular}[c]{@{}c@{}}Time to Compute \\ Gradients (seconds)\end{tabular}    & 41.2      & 28.1     & 12.9     & 10.5     \\\hline
\begin{tabular}[c]{@{}c@{}}Estimated EC2 \\ instance Cost \\ (USD / seconds )\end{tabular} & \$0.00000639 & \$0.00000639 & \$0.00000639 & \$0.00000639 \\\hline
\begin{tabular}[c]{@{}c@{}}Estimated Lambda \\ Cost (USD / seconds)\end{tabular} & \$0.0000573  & \$0.0000362  & \$0.0000233  & \$0.0000220  \\\hline
\begin{tabular}[c]{@{}c@{}}Estimated Compute \\ Gradients Cost per \\ Peer (USD)\end{tabular} & \$0.03567     & \$0.03069     & \$0.03451     & \$0.05435    \\ \hline
\end{tabular}
\end{adjustbox}
\label{tab:with_serverless}
\end{table}

\begin{table}[H]
\caption{Time and Cost Evaluation of Compute Gradients in Peer to Peer Training without Serverless; Model trained on VGG11, MNIST dataset, and Four Peers}
\centering
\begin{adjustbox}{max width=9cm}
\begin{tabular}{|c|c|c|c|c|}
\hline
batch size & 1024 & 512 & 128 & 64 \\ \hline
Instance Type & t2-large & t2-large & t2-large & t2-large \\ \hline
\begin{tabular}[c]{@{}c@{}}Time to Compute \\ Gradients  (seconds)\end{tabular} & 258 & 278,4 & 330,4 & 394,8 \\ \hline
\begin{tabular}[c]{@{}c@{}}Estimated EC2 \\ instance Cost \\ (USD / seconds )\end{tabular} & \$0.00002578 & \$0.00002578 & \$0.00002578 & \$0.00002578 \\ \hline
\begin{tabular}[c]{@{}c@{}}Estimated Compute \\ Gradients Cost per \\ Peer (USD)\end{tabular} & \$0.00665 & \$0.00717 & \$0.00851 & \$0.01017 \\ \hline
\end{tabular}
\end{adjustbox}
\label{tab:without_serverless}
\end{table}

\subsection{Compression and Communication Overhead}

In this section, we will explore Compression and Communication Overhead in distributed deep learning systems. We will first analyze the impact of varying the number of workers on computation and communication overhead, followed by an investigation into Gradient Compression techniques for enhancing communication efficiency during the training process.

\subsubsection{Computation and communication Over workers}
To elucidate the impact of communication overhead on system performance in a peer-to-peer architecture, we conducted rigorous experiments involving both VGG111 and MobileNet V3 Small models, varying the number of workers. In each experiment, we meticulously recorded both the compute time and the communication time. The results presented in the Figures \ref{fig:comm} show the relationship between the number of workers (peers), communication time, and computation time for VGG11 and MobileNet V3 Small models when using a batch size of 1024. In both cases, the figures reveal that as the number of workers increases, computation time decreases while communication time increases. This can be attributed to the fact that with more workers, the dataset is divided among more devices, allowing for faster computation. We notice that the magnitude of the increase is much higher in the VGG11 model compared to the MobileNet V3 Small model. This could be due to the VGG11 model having a larger number of parameters, which results in more gradient information being communicated between workers.

\subsubsection{Gradient Compression for communication improvement}
As mentioned in the previous section, communication overhead increases as the number of workers increases. Gradient compression can be a solution to mitigate this. In order to assess the impact of gradient compression on communication overhead, an experimental investigation was executed using the VGG11 model, the MNIST dataset, and a network composed of four peers. Our paramount focus was on precisely measuring the send and receive times from a single peer, in order to comprehensively elucidate communication efficiency. As illustrated in Figure\ref{fig:compression}, we demonstrate that the utilization of gradient compression techniques yields to a significant reduction in communication time when compared to the utilization of non-compressed gradients. This reduction in communication time is observed across a broad range of batch sizes. 
\begin{figure}[h]
 \centering
 \includegraphics[width=7.5cm]{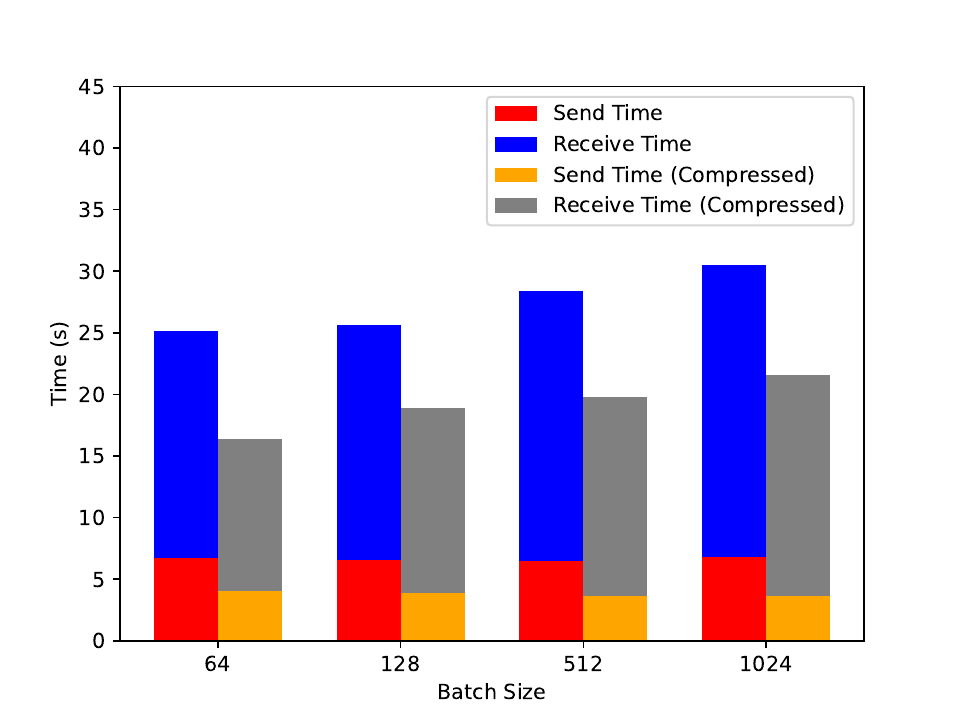}
 \caption{Compression Algorithm impact on Time Communication (Send and Receive Gradients)}
 \label{fig:compression}
\end{figure}

\subsection{Peer to Peer Training and Communication Barrier Synchronisation} 
In our experiments, we aimed to compare the performance of two different peer-to-peer (P2P) approaches: synchronous P2P and asynchronous P2P. We conducted experiments on Mobilenet v3 small with a batch size of 64, a learning rate of 0.001, and the optimizer SGD. Our findings revealed that the synchronous P2P approach outperformed the asynchronous P2P approach in terms of convergence rate and achieved a higher accuracy level. Specifically, the synchronous P2P approach achieved an accuracy of 84.3\% after approximately 128 epochs, while the asynchronous P2P approach required a greater number of epochs to converge and exhibited instability during the convergence. This was due to the asynchronous approach's tendency to consider outdated gradients, resulting in more epochs number to converge \cite{P46}. These results indicate that in P2P communication, synchronicity plays a crucial role in achieving a faster and more accurate convergence rate.

\begin{figure}[H]
 \centering
 \includegraphics[width=7.5cm]{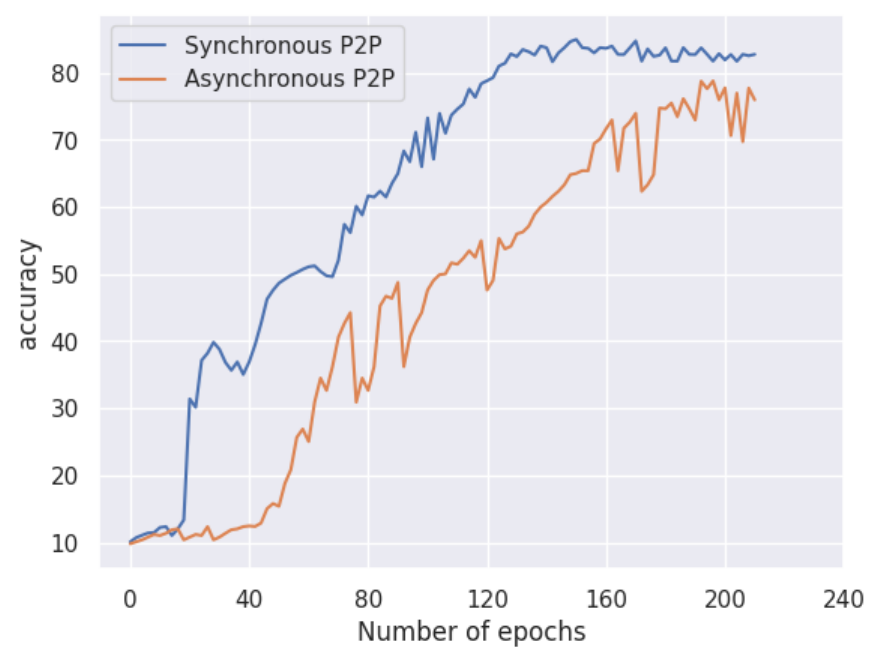}
 \caption{Synchronous Vs Asynchronous Peer to Peer Training of MobileNet V3 Small}
 \label{fig:loss_function}
\end{figure}

\section{Discussion}
\label{sec:discussion}

In this section, we reflect on the key findings and implications of our research on distributed deep learning in peer-to-peer training setups. Our focus is on the benefits and challenges of serverless infrastructure, the significance of managing communication overhead, addressing synchronization barriers, and the impact of choosing the model architecture and dataset on overall training performance.

\subsection{Benefits and Challenges of Serverless Infrastructure}

Our experimental results have underscored that the computation of gradients is the most resource-intensive task in distributed peer-to-peer training setups. By transitioning this task to serverless infrastructure, we have witnessed substantial improvements in computation time across various batch sizes and numbers of workers. This emphasizes the potential of serverless infrastructure as an efficient and scalable solution for managing computationally demanding tasks in distributed training, especially when working with complex models that require expensive computational instances.

Serverless infrastructure offers notable benefits but also presents cost implications. Our analysis revealed its potential in situations of constrained computational resources. Although serverless architectures may be more costly for smaller batch sizes, they significantly improve computation time when resources are scarce. This underlines their advantage in scenarios that demand rapid processing under resource limitations. Therefore, researchers and practitioners need to weigh their specific needs, such as time constraints, budget, resource limitations, and model complexity, when deciding on adopting a serverless architecture. This balanced approach allows for an optimal decision that accommodates cost, efficiency, and performance requirements.

\subsection{Reducing Communication Overhead}

Our analysis of communication overhead in distributed peer-to-peer training revealed that the amount of data transferred during the training process is influenced by the choice of model architecture and dataset. Larger model architectures and more complex datasets resulted in higher communication overhead, which can negatively impact the overall efficiency of the training process.

To reduce communication overhead, techniques such as gradient compression \cite{alistarh2017qsgd}, model sparsification \cite{ma2021effective}, Delta compression \cite{beznosikov2020biased}, communication optimisation \cite{zhou2021communication} and efficient data encoding can be employed. These methods can help minimize the data transfer during the training process, resulting in a more efficient and cost-effective training setup.

\subsection{Impact of Model Architecture and Dataset Choices}

Our experiments have shown that the choice of model architecture and dataset can have a significant impact on various aspects of distributed peer-to-peer training, including computational resource requirements, communication overhead, and synchronization barriers. Larger model architectures and more complex datasets generally require more computational resources, result in higher communication overhead, and demand longer synchronization times.

This highlights the importance of selecting appropriate model architectures and datasets for distributed training processes, considering the available resources and the desired trade-offs between training time, cost, and performance.
\section{Related work}
\label{sec:relatedwork}

In this related work section, we will explore two distinct but interrelated areas: Peer-to-Peer Machine Learning, which focuses on decentralized training approaches, and Serverless Computing for Machine Learning, which examines the efficient use of serverless to reduce computational overhead.

\subsection{Peer to Peer in Machine Learning} 
In recent years, numerous initiatives have been proposed to address the challenges of distributed, decentralized, and peer-to-peer (P2P) systems for machine learning. These works can be broadly classified into the following categories: decentralized training methodologies, privacy-preserving approaches, and communication-efficient solutions.

Decentralized training methodologies, such as BrainTorrent \cite{roy2019braintorrent} and the consensus-based distributed stochastic gradient descent algorithm proposed by Zhanhong et al. \cite{jiang2017collaborative}, utilize fully decentralized systems for training a shared model. Both approaches showcase the potential of decentralized training in large-scale machine learning systems while maintaining scalability and convergence guarantees. BrainTorrent requires peers to share their local model weights and update them by calculating a weighted average between the weights of the receiving and sending peers. In contrast, Zhanhong et al.'s algorithm leverages gossip-based communication protocols to propagate the model between nodes in fixed topology networks.

Addressing communication overhead and efficiency in P2P topologies is another critical aspect of distributed machine learning systems. Garfield \cite{guerraoui2021garfield} presents a decentralized architecture for training machine learning models in the presence of adversarial nodes, leveraging a Byzantine fault-tolerant consensus protocol for secure and scalable P2P training. Xing et al. \cite{xing2016strategies} highlight the communication overhead in P2P parameter synchronization and the need for efficient communication strategies. SELMCAST \cite{luo2022fast}, an algorithm for multicast receiver selection, optimizes the bottleneck sending rate to reduce time cost for parameter synchronization. Lastly, the SAPS-PSGD algorithm \cite{tang2020communication} maximizes bandwidth efficiency through adaptive worker pair selection in a distributed training approach involving a coordinator and multiple peers. These communication-efficient solutions showcase the potential to overcome the challenges of communication overhead in peer-to-peer topologies while achieving substantial savings compared to centralized topologies.

In this work, our proposed method involves peers in a distributed training system sending large gradient computations to serverless computing resources instead of calculating them locally. By doing so, the peers can focus on other tasks, such as updating model weights and communicating with other nodes in the network, while serverless platforms efficiently handle the computationally expensive gradient calculations.

\subsection{Serverless Computing for Machine Learning}

Given that the use of serverless runtimes for machine learning pipelines is a relatively new research area \cite{barrak, P08, P15, P16, P34}, several initiatives have been undertaken to encourage wider adoption and promote efficient utilization of Function-as-a-Service (FaaS) platforms. To this end, various serverless development frameworks have been proposed, such as Cirrus \cite{P08}, which has been meticulously crafted to proficiently manage the entire ML workflow.

Particularly, recent research efforts have been directed towards ML model training \cite{P51, P53, P49, P46, P39, P38}. Existing approaches for distributed training machine learning are based on parameter server communication topologie where all communication between workers goes through a server for a synchronisation purpose.

Ali et al. \cite{P54} proposed SMLT, a serverless framework for distributed training based on parameter server architecture and a Hybrid Storage Enabled Hierarchical Model Synchronization method. This approach achieves faster training speeds and reduced monetary costs compared to other serverless ML training frameworks and VM-based systems. Experimental evaluations show SMLT outperforms state-of-the-art VM-based systems and serverless ML training frameworks in training speed (up to 8×) and cost (up to 3×).

MLLESS \cite{P53}, proposed by Sarroca and Sanchez-Artigas, is a FaaS-based ML training prototype designed for cost-effective ML training in serverless computing. It incorporates a decentralized design, a significance filter, and a scale-in auto-tuner optimized for serverless computing. MLLess outperforms serverful ML systems by up to 15x for sparse ML models with fast convergence and demonstrates ease of scaling out to large clusters of serverless workers.

In this work, we use a fully decentralized machine learning training approach, leveraging serverless benefits to compute expensive gradient calculations, combining the advantages of both distributed learning and serverless computing for more efficient and scalable ML training processes.

\section{Conclusion}
\label{sec:conclusion}
In this paper, we present a novel serverless peer-to-peer (P2P) architecture for distributed training, introducing an efficient parallel gradient computation technique to address resource constraints. We evaluated the performance of our approach with a focus on computational resource requirements, serverless infrastructure efficiency, communication overhead, and synchronization barriers. Our experimental results showed that the computation of gradients is the most computationally expensive task, benefiting from serverless infrastructure integration and leading to up to a 97.34\% improvement in computation time. We investigated the trade-off between computation time improvements and associated costs, revealing that serverless architecture tended to be more expensive, with costs being up to 5.3 times higher than traditional, instance-based architectures. Additionally, we briefly analyzed the communication overhead and synchronization problems in the distributed training process, highlighting the need for efficient strategies in P2P distributed training systems.

The insights gleaned from this research can be utilized by other researchers and practitioners to build upon our work, further optimize distributed training processes, and potentially revolutionize the way machine learning models are trained across various applications.

\bibliographystyle{IEEEtran}

\bibliography{bibliography}
\end{document}